\documentclass[reprint,superscriptaddress,amsmath,amssymb,aps,pre]{revtex4-1}

\usepackage{amsfonts}
\usepackage{amsmath,amssymb}
\usepackage{graphicx}
\usepackage{dcolumn}
\usepackage{bm}
\usepackage{color}
\usepackage{bbold}
\usepackage{soul}
\usepackage{placeins}
\usepackage{epsfig,epic}
\newcommand{\newc}{\newcommand}
\newc{\FIXME}[1]{{\color{red}FIXME: #1}}
\newc{\TODO}[1]{{\color{green}TODO: #1}}
\newc{\beq}{\begin{equation}}
\newc{\eeq}{\end{equation}}
\newc{\kt}{\rangle}
\newc{\br}{\langle}
\newc{\beqa}{\begin{eqnarray}}
\newc{\eeqa}{\end{eqnarray}}
\newc{\longra}{\longrightarrow}

\providecommand*{\Pab}{P_{a:b}}
\providecommand*{\Mab}{M_{a:b}}
\providecommand*{\Vab}{V_{a:b}}

\providecommand*{\Aab}{\mathcal{A}_{a:b}}
\providecommand*{\kml}{k_{m,l}}
\providecommand*{\Pml}{P_{m,l}}
\providecommand*{\Prat}{P_{\mathrm{rat}}}
\providecommand*{\Pmlt}{\tilde{P}_{m,l}}
\providecommand*{\Pratt}{\tilde{P}_{\mathrm{rat}}}
\providecommand*{\Rekml}{\mathrm{Re}\,k_{m,l}}
\providecommand*{\ReRkml}{\mathrm{Re}\,k_{m,l}R}
\providecommand*{\Imkml}{\mathrm{Im}\,k_{m,l}}
\providecommand*{\ImRkml}{\mathrm{Im}\,k_{m,l}R}

\providecommand*{\dpos}{\Delta p_{\mathrm{POS}}}
\providecommand*{\dPos}{\Delta P_{\mathrm{POS}}}
\providecommand*{\AT}{\mathcal{A}_{T}}

\makeatletter
\let\Hy@backout\@gobble
\makeatother

\begin{document}

\title{Resonance--Assisted Tunneling in Deformed Optical Microdisks \\
	with a Mixed Phase Space}

\author{Felix Fritzsch}
\affiliation{Technische Universit\"at Dresden, Institut f\"ur Theoretische
             Physik and Center for Dynamics, 01062 Dresden, Germany}

\author{Roland Ketzmerick}
\affiliation{Technische Universit\"at Dresden, Institut f\"ur Theoretische
	Physik and Center for Dynamics, 01062 Dresden, Germany}
\affiliation{Max-Planck-Institut f\"ur Physik komplexer Systeme, N\"othnitzer
	Stra\ss{}e 38, 01187 Dresden, Germany}

\author{Arnd B\"acker}
\affiliation{Technische Universit\"at Dresden, Institut f\"ur Theoretische
             Physik and Center for Dynamics, 01062 Dresden, Germany}
\affiliation{Max-Planck-Institut f\"ur Physik komplexer Systeme, N\"othnitzer
Stra\ss{}e 38, 01187 Dresden, Germany}

\date{\today}

\begin{abstract}
	The life times of optical modes in whispering-gallery cavities crucially
	depend on the underlying
	classical ray dynamics and may be spoiled by the presence of classical
	nonlinear resonances due to resonance--assisted tunneling.
	Here we present an intuitive semiclassical picture which allows for
	an accurate prediction of decay rates of optical modes
	in systems with a mixed phase space.
	We also extend the perturbative description from near-integrable systems to
	systems with a mixed phase space and find equally good agreement.
	Both approaches are based on the approximation of the actual ray dynamics by
	an integrable Hamiltonian, which enables us to perform a semiclassical
	quantization of the system and to introduce a ray-based description of the
	decay of optical modes.
	The coupling between them is determined either perturbatively or semiclassically
	in terms of complex paths.
\end{abstract}

\pacs{PACS here}

\maketitle

\section{Introduction}

Optical microcavities allow for a wide range of applications
\cite{Vah2004, CaoWie2015} in, for instance, sensors \cite{ForSwaVol2015}
and lasing devices \cite{HeOezYan2013}.
Optical modes
with long life times and directional emission properties are
desired.
Long life times, and thus also large quality factors, are realized by
whispering--gallery cavities in which light
is confined by almost total internal reflection.
This is achieved by a circular or spherical cavity design
for which the classical ray dynamics is integrable.
In contrast, directional emission is achieved by deforming the cavities
boundary from the perfectly circular or spherical shape
and thus rendering the classical ray dynamics nonintegrable
\cite{NoeStoCheGroCha1996, WieHen2008}.
In quasi two-dimensional cavities classical
whispering--gallery
trajectories persist under sufficiently small and
smooth deformations.
This, in general, gives rise to the coexistence
of regular and chaotic ray dynamics in a mixed phase-space \cite{Laz1973a}.
While the associated whispering--gallery modes are still present in the
deformed cavity, their quality factors get spoiled, i.e., their life times
decrease.

One major mechanism causing this enhanced decay
is the wave effect of dynamical tunneling which was first observed and
extensively studied in quantum systems \cite{DavHel1981, KesSch2011}.
More recently it has been studied both theoretically and experimentally
in microwave resonators \cite{DemGraHeiHofRehRic2000,
BaeKetLoeRobVidHoeKuhSto2008,DieGuhGutMisRic2014, GehLoeShiBaeKetKuhSto2015}
and optical microcavities \cite{HacNoe1997, PodNar2005,BaeKetLoeWieHen2009,
	 ShiHarFukHenSasNar2010,
ShiHarFukHenSunNar2011,YanLeeMooLeeKimDaoLeeAn2010, KwaShiMooLeeYanAn2015,
YiYuLeeKim2015, YiYuKim2016, ShiSchHenWie2016, KulWie2016b, YiKulKimWie2017}.
Dynamical tunneling allows for coupling of optical modes associated with
dynamically separated classical phase-space regions.
Specifically, long living whispering--gallery modes may couple to faster
decaying modes via dynamical tunneling.
In particular, classical nonlinear resonances in the ray dynamics may
drastically enhance
tunneling effects by the mechanism of resonance--assisted tunneling
\cite{BroSchUll2002}.
Again the majority of theoretical understanding has been obtained
in quantum systems, e.g., quantum maps
\cite{BroSchUll2001, BroSchUll2002, EltSch2005, LoeBaeKetSch2010,
	SchMouUll2011}.
Recent experiments, however, have impressively demonstrated the effect also
in microwave resonators \cite{GehLoeShiBaeKetKuhSto2015} and in
two-dimensional optical microcavities \cite{KwaShiMooLeeYanAn2015}.
While the
latter qualitatively follows the theoretical description of
resonance--assisted tunneling obtained in quantum maps a quantitative
description of resonance assisted tunneling in optical microcavities has been
given only recently for systems with near-integrable ray dynamics
\cite{KulWie2016b, YiKulKimWie2017}.
For this, the classical ray dynamics has been approximated by a
pendulum-like Hamiltonian, which subsequently allows for the perturbative
expansion of optical modes and their quality factors as predicted by
resonance--assisted tunneling.
Moreover, this description has been shown to capture the enhanced decay of
optical modes correctly in situations when the perturbative scheme developed in
Ref.~\cite{DubBogDjeLebSch2008} fails.
Nevertheless, the prediction based on resonance--assisted tunneling has
two major limitations.
On the one hand it depends on the circular cavity yielding a good approximation
of the actual ray dynamics in order to construct the approximating
pendulum Hamiltonian.
This will not be accurate for larger deformations of the cavitiy boundary.
On the other hand for larger deformations also the optical modes of the
circular cavity will no longer provide a suitable unperturbed basis for the
perturbative treatment.

In this paper we extend the
perturbative description of resonance--assisted tunneling in optical
microcavities to systems with a mixed phase space.
In addition, we apply the semiclassical description of resonance--assisted
tunneling which leads to a simple and intuitive picture.
To this end, based on an idea presented in Ref.~\cite{RobBer1985}, we introduce
a suitable coordinate system and subsequently construct the
approximating pendulum Hamiltonian.
These coordinates further allow for semiclassical Einstein--Brillouin--Keller
(EBK) quantization of the
system \cite{KelRub1960, NoeSto1997, Noe1997}.
This enables us to compute wave numbers as well as their associated
classical phase-space structures.
Moreover, this provides the unperturbed basis necessary for the perturbative
treatment and thus allows for the construction of
couplings between modes due to resonance--assisted tunneling.
We further use both the approximating pendulum Hamiltonian and the
semiclassical EBK quantization scheme to obtain a semiclassical description of
these couplings.
Combining the coupling of different modes with a ray-based model for
their decay finally allows to accurately describe the complex wave numbers and
thus the decay of optical modes in the presence of
resonance--assisted tunneling.
Both the semiclassical and perturbative description agree well with numerical
solutions of the mode equation.

This paper is organized as follows:
In Sec.~\ref{sec:microdisks} we introduce deformed optical microdisks.
We study their classical ray dynamics in Sec.~\ref{sec:ray_dynamics}
and introduce the full wave picture in terms of the mode equation in
Sec.~\ref{sec:mode_eq}.
In Sec.~\ref{sec:numerics} we discuss numerical solutions to the mode equations
and briefly review a perturbative approach and its applicability to
resonance--assisted tunneling in Sec.~\ref{sec:perturbation_th}.
Sec.~\ref{sec:RAT} introduces the basic concepts of our description.
This includes the introduction of adiabatic action angle coordinates in
Sec.~\ref{sec:CanonicalTrafo} and the subsequent construction of an
approximating pendulum Hamiltonian in Sec.~\ref{sec:pend_ham}.
The semiclassical quantization scheme is discussed in Sec.~\ref{sec:ebk}
and a ray-based model for the decay of optical modes is introduced in
Sec.~\ref{sec:ray_decay}.
Finally, Sec.~\ref{sec:rat_pert_th} contains the perturbative description of
resonance--assisted tunneling while the semiclassical picture is presented in
Sec.~\ref{sec:RAT_sc}.
A summary and outlook are given in Sec.~\ref{sec:sum_outlook}.

\section{Deformed Optical Microdisks \label{sec:microdisks}}

In this section we introduce deformed optical microdisks and present the model
systems studied in this paper.
We discuss their classical ray dynamics as well as solving Maxwells equations to
obtain the optical modes and their decay rates.

\subsection{Ray Dynamics \label{sec:ray_dynamics}}

\begin{figure}[]
\includegraphics[width=8.6cm]{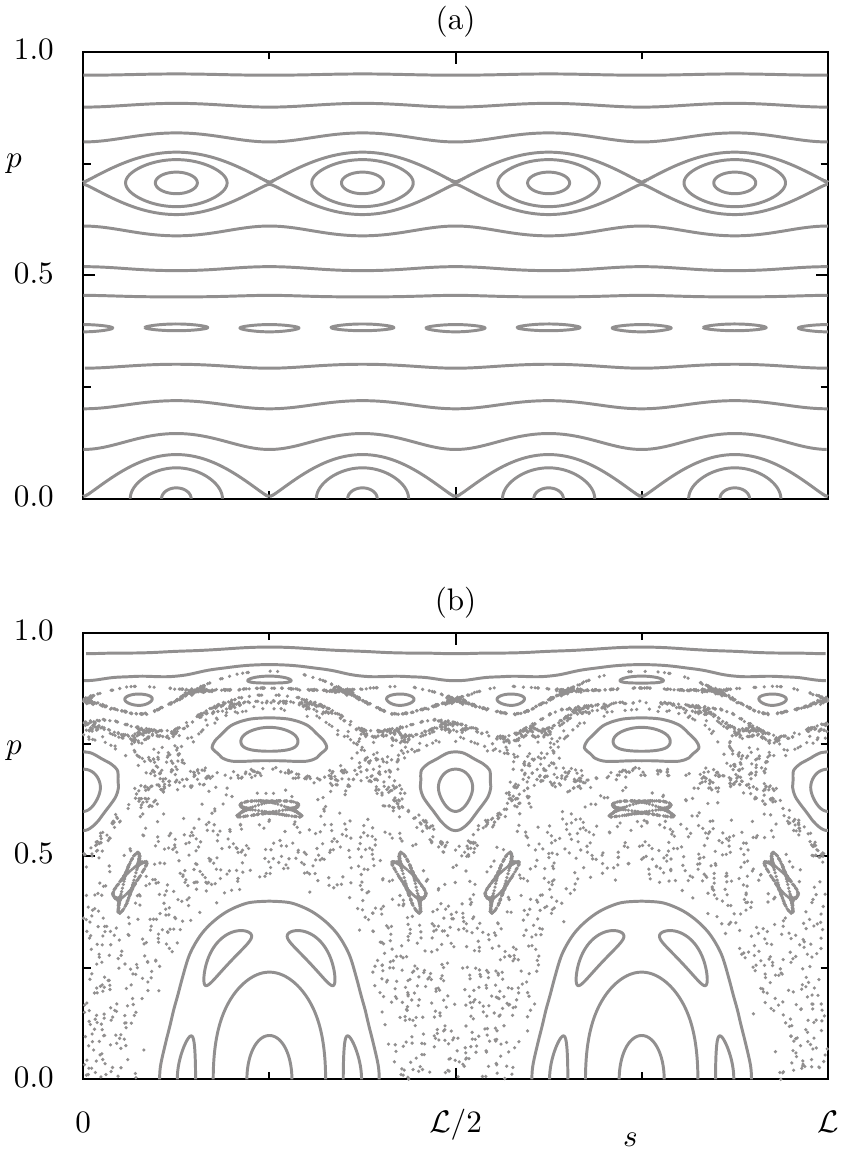}
\caption{\label{fig:phase_space} (a) Phase space of the
         near-integrable system and
	(b) phase space of the mixed
	system with parameters given in the text. Regular tori are depicted by lines, while dots correspond to
	chaotic orbits.}
\end{figure}

The classical ray dynamics of a cavity is given by light rays traveling on
straight lines inside the cavity
and specular reflections at the boundary.
Thus the dynamics is completely determined by the positions of the bounces on
the boundary and the angle  $\chi$ of the reflected ray with the inwards normal
vector of the boundary.
Therefore the dynamics can be completely described in Birkhoff
coordinates $(s, p)$, where $s$ denotes the arc length along the boundary
ranging from $0$ to the full boundary length $\mathcal{L}$ and $p=\sin(\chi)$
is the canonically conjugate momentum.
In Birkhoff coordinates, the dynamics is conveniently represented in the
phase space $(s, p) \in \left[0, \mathcal{L}\right]\times\left[-1, 1\right]$
by a symplectic map $(s, p) \mapsto (s^{\prime}, p^{\prime})$ given by
successive bounces and the corresponding momenta.

Typically, the dynamics generated by this map will be nonintegrable, as
the full system has two degrees of freedom while there is no second conserved
quantity besides energy.
In fact, there are only a few cavities with integrable ray dynamics and smooth
boundary.
Among them, the circular cavity is the most simple case.
However, a generic deformation of a circular boundary will render the ray
dynamics nonintegrable.
Typically, with growing deformation the dynamics will change from near
integrable for small deformations towards a mixed phase space for larger
deformations.
For even larger deformations the dynamics can become fully chaotic.
An important exception is the elliptic cavity, in which
the eccentricity can be seen as deformation parameter.
There, the ray dynamics remains integrable
independent of the eccentricity and thus even for large deformations.
The boundary of a deformed circular cavity can be described
by its radial coordinate $r$ as a function of the polar angle $\varphi$.
For cavities with reflection symmetry with respect to an axis and with
smooth boundaries this function can be written as a Fourier series,
$r(\varphi) = R\left(1 + \sum_{j \geq 1}d_j \cos(j \varphi)\right)$,
with mean radius $R$.
As concrete examples we will consider cases where only one of the coefficients
$d_N:=\epsilon$ is nonzero, i.e.,
\begin{align}
r(\varphi) = R\left(1 + \epsilon \cos(N\varphi)\right).
\label{eq:boundary}
\end{align}

In particular we choose $R=1$, $\epsilon = 0.0025$ and $N=4$ as an example of a
weakly
deformed cavity with near-integrable ray dynamics which was also considered in
Ref.~\cite{KulWie2016b}.
Its phase space, shown in Fig.~\ref{fig:phase_space}(a), is
predominantly foliated by invariant tori.
At $p=0$ there are two stable and two unstable period-two orbits, which
correspond to trajectories
along the diameter of the cavity.
In phase space this periodic motion leads to orbits consisting of only
two points each.
It can be characterized by a frequency
$\omega = \mathcal{L}/2$, by which the arc-length coordinate advances
in between successive bounces.
The periodicity of these orbits is reflected by the fact that this frequency
fulfills a resonance condition, i.e., it is a
rational multiple of $\mathcal{L}$.
Therefore one has
\begin{align}
\omega = \omega_{a:b} = b\mathcal{L}/a
\label{eq:resonance_condition}
\end{align}
for integers $a$ and $b$.
The period of the orbits is given by the numerator of $a/b$ after fully
reducing this fraction.
In phase space this gives rise to a chain of $a$ eye-like structures, called a
nonlinear resonance.
The frequency $\omega_{a:b}$ is such, that the central periodic orbits
within these resonance eyes advance by $b$ eyes in between successive bounces.
Thus, for the two pairs of stable and unstable period-two orbits, a
$a:b=4:2$ resonance chain arises in phase space.
It is bounded by a very thin chaotic layer,
which can not be distinguished from regular tori at the shown scale.
In addition a large
$4:1$ resonance occurs around $p_{4:1}=\cos(\pi/4) = 1/\sqrt{2}$.
For larger momenta $p \to 1$ again motion along regular tori dominates the
phase-space portrait.
In configuration space the corresponding trajectories closely follow the
boundary of the cavity
and are therefore called whispering--gallery trajectories.
Their existence is guaranteed by a theorem of Lazutkin \cite{Laz1973a} if the
curvature of a smooth boundary is bounded from below and above by some positive
constants.
As the deformation parameter $\epsilon$ is small the shape of the boundary does
not deviate much from the circular cavity.
Thus the regular dynamics of the circular cavity approximates the motion along
regular tori of the weakly deformed system well. \\

Another example we consider is given by  $R=1$, $\epsilon = 0.08$ and $N=2$.
This deformation is called
quadrupol deformation and, although the deformation parameter $\epsilon$ is
again small  compared to unity, it gives rise to a mixed phase space as it
is shown in Fig.~\ref{fig:phase_space}(b).
Hence, in the following we will use the shorthand notion of a mixed system for
this example.
For small momenta the phase space is governed by the regular islands
around the stable period-two orbit and the
chaotic regions around the unstable period two-orbit at $p=0$.
For large momenta, $|p|$ close to $1$, again Lazutkins
theorem implies the existence
of regular motion along invariant tori which correspond to whispering--gallery
trajectories in configuration space.
Below the last invariant torus a large $4:1$ resonance occurs around
$p_{4:1}=1/\sqrt{2}$.
More precisely, the unstable period-four orbit has constant momentum $p_{4:1}$,
while its stable counterpart oscillates around this value.
This $4:1$ nonlinear resonance chain is surrounded by a partial barrier
as manifested by different densities of points in the chaotic region in
Fig.~\ref{fig:phase_space}(b).
Inside the chaotic regions smaller nonlinear resonances of order proportional
to $N=2$ can be found.

The quadrupol deformation agrees in lowest order of the deformation parameter
with the elliptical cavity.
That is, the boundary of the elliptic cavity with eccentricity $\epsilon_{\mathrm{ecc}}$ in polar coordinates can be written as
\begin{align}
r(\varphi)=\frac{R\sqrt[4]{1 - \epsilon_{\mathrm{ecc}}^2}}{\sqrt{1 - \epsilon_{\mathrm{ecc}}^2\cos^2(\varphi)}},
\end{align}
which for small $\epsilon_{\mathrm{ecc}}$ can be expanded as \cite{YiKulKimWie2017}
\begin{align}
r(\varphi)\approx R\left(1 + \epsilon\cos(2\varphi)\right).
\end{align}
Here $\epsilon$ and the eccentricity $\epsilon_{\mathrm{ecc}}$ are related via
\begin{align}
\epsilon_{\mathrm{ecc}} = \sqrt{1 - (1 - \epsilon)^4}.
\label{eq:eccentricity}
\end{align}
Thus the motion along regular tori corresponding to whispering-gallery
trajectories in the mixed system can be approximated by
the regular dynamics of the elliptic cavity with eccentricity given by
Eq.~\eqref{eq:eccentricity}. \\

As the billiard-like dynamics inside a cavity corresponds to the
classical ray picture of optics it is expected to be valid only in the limit of
vanishing wave length or equivalently large wave numbers $k$.
For smaller wave numbers semiclassical corrections to the classical ray picture
may become relevant.
Most famous among these corrections is the so-called Goos--H{\"a}nchen shift
\cite{GooHae1947, Art1948}.
It describes the deplacement between the center of an incoming light beam and
the center of the reflected beam.
In particular, this causes the position of periodic orbits and the associated
nonlinear resonance chain to undergo a shift in momentum \cite{UntWie2010}.
This periodic-orbit shift
is given by \cite{UntWie2010, KulWie2016b}
\begin{align}
\dpos(k) = \frac{p_{a:b}}{R_{\mathrm{c}}\mathrm{Re\,}k\sqrt{n^2p_{a:b}^2 - 1}},
\label{eq:dpos}
\end{align}
where $p_{a:b}=\cos(b\pi/a)$ for an $a:b$ resonance.
Here, $n$  denotes the refractive index inside the
cavity and $R_{\mathrm{c}}$ is the average of the radii of curvature of the
boundary
taken over all points of the stable and unstable periodic orbit.
Equation~\eqref{eq:dpos} will be used in the following, when comparing
classical
phase-space structures with solutions of Maxwells equations.

\subsection{Mode Equation \label{sec:mode_eq}}

While the classical dynamics inside the cavity is governed by light rays
traveling along straight lines and undergoing specular reflections at the
boundary,
the electro--magnetic field
is described by Maxwells equations in full space.
Solutions to Maxwells equations with harmonic time dependence $\propto
e^{\mathrm{i}\omega t}$
in a quasi two-dimensional cavity fulfill the mode equation
\begin{align}
\Delta \psi(\mathbf{r}) + n({\mathbf{r}})^2k^2\psi({\mathbf{r}})=0
\label{eq:mode_eq}
\end{align}
in combination with appropriate boundary conditions.
Here, $k=\omega/c$ denotes the wave number while $c$ is the speed of light in
vacuum.
The effective refractive index $n({\mathbf{r}})$ in Eq.~\eqref{eq:mode_eq} is
considered
to be constant, $n({\mathbf{r}}) = n$, inside the cavity and is set to unity on
the outside in the following.
For numerical calculations we use $n=2$.
Depending on the polarization the optical mode $\psi(\mathbf{r})$ corresponds
to either the
component perpendicular to the cavity plane of the
electric field, which is called transversal magnetic (TM)
polarization, or of the magnetic field, called transversal electric (TE) polarization.
For TM polarization $\psi$ and its normal derivatives are continuous at the cavities boundary.
While we will focus on TM polarized light, we expect the results of this paper
to be
applicable also in the TE case when the appropriate boundary conditions are
applied.
Requiring also outgoing wave boundary conditions at infinity, i.e.,
an asymptotic behavior $\text{e}^{\mathrm{i}kr}/\sqrt{r}$ for large $r$,
Eq.~\eqref{eq:mode_eq} has solutions only for discrete complex wave
numbers with negative imaginary part.
If the boundary of the cavity exhibits reflection symmetry along an axis, the
optical modes reflect that symmetry, i.e., they fulfill either Dirichlet or
Neumann boundary conditions on that symmetry axis.
Thus in the cavities under consideration, which are symmetric with respect to
both horizontal and vertical axis, the optical modes are grouped into four
symmetry classes.

By means of semiclassical quantization in classically integrable or mixed
systems at least some of
the possible modes and their mode numbers $k=k_{ml}$ are labeled by an angular
mode number $m$ and a radial mode number $l$.
In particular modes with small radial mode number $l$ are associated with
whispering--gallery
trajectories and thus are called whispering--gallery modes.
As the wave numbers are complex the intensity of optical mode inside the cavity
decays as
$\text{e}^{-\Gamma_{m,l}t}$ with a decay rate $\Gamma_{m,l}= -2c \, \Imkml$, which is
exponentially small for whispering--gallery modes and thus corresponds to large
life times $\tau_{m,l} = 1/\Gamma_{m,l}$.
This is often also quantified by the quality factor
$Q_{m,l} = - \Rekml/(2 \Imkml)$.
As for both decay rates and for quality factors it is sufficient to determine
the wave numbers $\kml$ we will focus on these in the
following.
In particular, we study whispering--gallery modes with fixed radial
quantum number $l=0$.

\begin{figure}[t]
	\includegraphics[width=8.6cm]{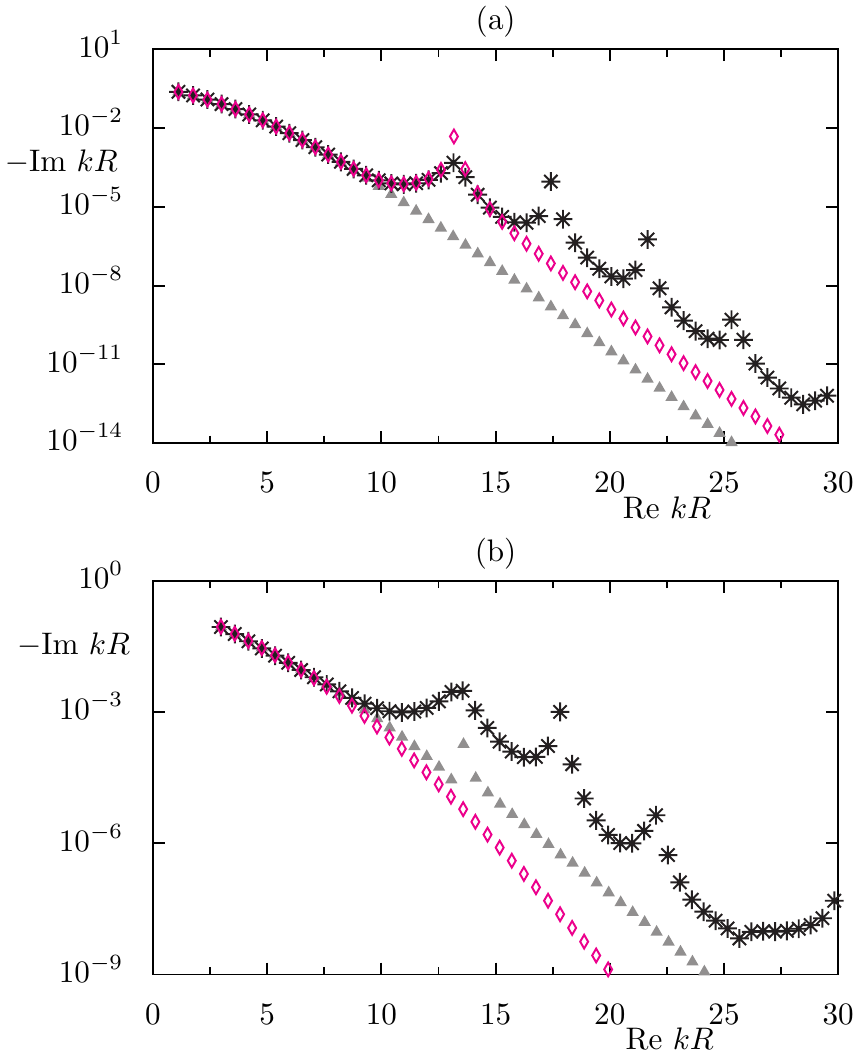}
	\caption{\label{fig:wave_numbers0} Wave numbers for whispering--gallery
		modes with $l=0$ (a) for the near-integrable system starting from
		$m=1$
		and  (b) for the mixed system starting from $m=4$ as black stars.
		The wave numbers of the integrable counterpart, i.e., (a) the circular cavity and (b) the elliptic cavity are shown as gray triangles.
		The perturbative prediction, Eq.~\eqref{eq:boundary_PT}, is depicted by open magenta diamonds.}
\end{figure}

\subsection{Numerical Results \label{sec:numerics}}

An analytical solution to Eq.~\eqref{eq:mode_eq} and for the wave numbers
$\kml$ is only known
for the circular cavity.
Therefore in general numerical or approximate schemes have to be used.
A standard numerical approach both for closed billiards and cavities is
the boundary element method (BEM) \cite{Bae2003, Wie2003}, which we combine
with the ideas of Ref.~\cite{Kre1991} to gain the required accuracy for the
exponentially small imaginary wave numbers.
We use this approach to
determine the dimensionless wave numbers $R\kml$ for whispering gallery
modes with mode numbers $(m, l=0)$ in the
near-integrable system and the mixed system.
They are shown as black stars in Fig.~\ref{fig:wave_numbers0} for the (a)
near-integrable and (b) mixed system.
There, $-\ImRkml$ is
depicted semilogarithmically as a function of $\ReRkml$.
This representation is usually employed for decay rates in quantum maps plotted
against the
inverse semiclassical parameter, whose role is played by $\ReRkml$ in our case,
and is well suited
to study resonance--assisted tunneling.
We focus on modes which fulfill Dirichlet boundary conditions on the
horizontal symmetry axis.
Modes with even angular mode number $m$ additionally fulfill Dirichlet boundary
conditions
along the vertical symmetry axis, while for odd $m$ they fulfill Neumann boundary conditions.
The wave numbers of modes belonging to other symmetry classes show
qualitatively the same behavior and cannot be distinguished from the depicted
data on the shown scale.
For small $\ReRkml$ the imaginary parts follow an exponential decay, which is
equal to the wave numbers of its integrable counterpart, i.e., the
circular and elliptic cavity, respectively.
Their wave numbers are shown as gray triangles and are computed analytically
for the circular cavity and numerically using the BEM for the elliptic cavity.
Note, that for the elliptic cavity the imaginary parts of the wave numbers do
not show a
pure exponential decay as there are deviations around $\ReRkml = 13$.
This is expected to be due to the open boundary condition of optical cavities
\cite{YiKulKimWie2017}.
However, for both the near-integrable and the mixed system, at some point the
imaginary parts of the wave numbers deviate from the
integrable case due to resonance--assisted tunneling induced by the $4:1$
resonance.
There is still an overall exponential decay which is similar to the exponential
decay observed in the integrable case.
This overall decay is accompanied by peaks, at which the negative
imaginary part of the wave
number is enhanced by roughly two orders of magnitude.
While showing qualitatively similar behavior, the overall exponential decay is
slower in the case of the mixed system.
In the mixed system the overall exponential decay and the peak structure
are present only up to wave numbers with $\ReRkml = 25$.
For larger real parts of the wave numbers their imaginary parts start to form
a plateau and even increase towards $\ReRkml = 30$.
Similar effects have been observed in the studies on resonance--assisted
tunneling in quantum maps if additional smaller resonances
become important \cite{LoeBaeKetSch2010}.
In the near-integrable system this plateau formation is not observed in the investigated regime of wave numbers. \\

\begin{figure}[h]
	\includegraphics[width=8.6cm]{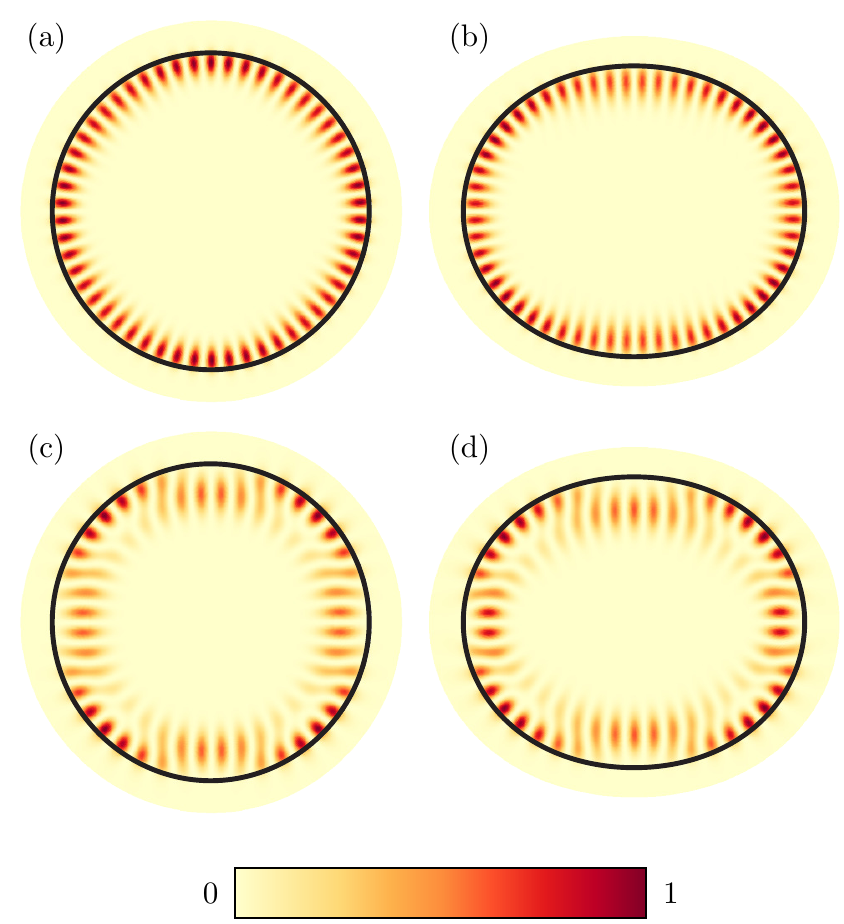}
	\caption{\label{fig:wave_functions}
		The intensity patterns with the respective maximum intensity normalized
		to unity
		for the modes with (a) $m=27$ and (c) $m=22$ for the near-integrable
		systems and
		(b) $m=28$ and (d) $m=23$ for the mixed system are shown. The black line
		represents the boundary of the respective cavity.}
\end{figure}

A qualitative idea of the mechanism causing the peaks for specific values of
$\ReRkml$
can be obtained from the intensity $|\psi(\mathbf{r})|^2$ of the optical modes
shown in
Fig.~\ref{fig:wave_functions}.
Here the intensity is shown for the mode with $m=27$ for the near-integrable system in
Fig.~\ref{fig:wave_functions}(a) and with $m=28$ for the mixed
system in (b) which corresponds to modes in between the
first two peaks.
In both cases the region of largest intensity basically follows the boundary of
the cavity
resembling classical whispering--gallery trajectories.
Closer inspection reveals that the shape of the rectangular stable period-four
orbit also
influences the morphology of the intensity patterns.
In contrast, the modes corresponding to the first peak are shown in
Fig.~\ref{fig:wave_functions}(c) for the near-integrable system with $m=22$ and
in (d) for the mixed system with $m=23$.
Both appear as a superposition of modes with different radial mode numbers.
In the near-integrable system
the admixture of modes with larger radial mode number $l>0$ is well seen in the
regions where the stable period
four orbit hits the boundary.
In the mixed system a similar pattern emerges where the unstable period-four
orbit hits the boundary.\\

\begin{figure}[h]
	\includegraphics[width=8.6cm]{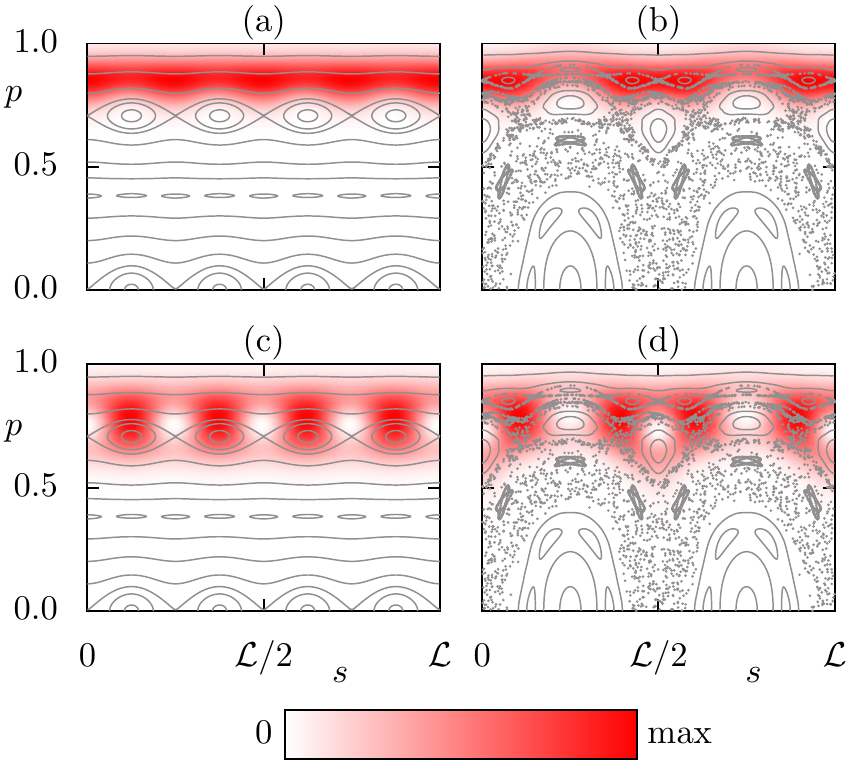}
	\caption{\label{fig:wave_husimis}
		The incident Husimi representation of the modes with (a) $m=27$ and (c) $m=22$ for the near-integrable
		systems and
		(b) $m=28$ and (d) $m=23$ for the mixed system are superimposed on the classical phase space shown in gray.}
\end{figure}

The above observations are also reflected in the phase-space representation of
the modes.
To this end we also show the incident Husimi function \cite{HenSchSch2003} on
the inside of the boundary in Fig.~\ref{fig:wave_functions} superimposed on the
classical phase space.
The Husimi functions are obtained from the overlap of $\psi$ and its normal
derivative with Gaussian coherent states defined on the boundary.
We choose the ratio of their uncertainties in $s$ and $p$ to be equal to the
boundary length $\mathcal{L}$ in order to increase the
resolution in $p$ direction and to take the extent of phase space in both $s$
and $p$ into account.
Doing so we find the mode with $m=27$ in the near-integrable system
to localize well above the $4:1$ resonance and to match the shape of the
regular tori as
shown in Fig.~\ref{fig:wave_husimis}(a).
Similarly, in the mixed system the mode with $m=28$ shown in (b) localizes 
above the
$4:1$ resonance where it follows the shape of adiabatically invariant curves, 
see Fig.~\ref{fig:adiabatic_invariants}(a).

The localization properties of the modes with $m=22$ in the near-integrable
system and with $m=23$ in the mixed system shown in
Fig.~\ref{fig:wave_husimis}(c)~and~(d), respectively, exhibit a different
morphology.
They predominantly localize on the $s$-coordinates of either the stable period
four orbit in the near-integrable system or of the unstable period-four orbit in
the mixed system.
However, in $p$ direction the modes are shifted upwards compared to the $p$
coordinates of the respective period-four orbits.
This is due to the periodic-orbit shift, Eq.~\eqref{eq:dpos}.
Incorporating this shift would lead to the modes localizing predominantly on
the stable periodic
orbit in the near-integrable system and on the unstable periodic orbit in the
mixed system, respectively (not shown).
Besides the regions of maximal intensity along the periodic orbits there are
additional phase-space regions with significant intensity.
This is most clearly seen in Fig.~\ref{fig:wave_husimis}(c) for the
near-integrable system.
In this case these regions correspond to two regular tori located
above and below the, according to Eq.~\eqref{eq:dpos} appropriately shifted,
$4:1$ resonance.
Taking again the periodic-orbit shift into account would result in those tori
being located symmetrical with respect to the resonance.
Thus the mode with $m=23$ may be interpreted as a superposition of modes
localizing on these two different classical tori.
In the mixed system the regions with additional intensity are located on either 
side of the $4:1$ resonance as well.
Above the resonance chain this contribution shows a similar morphology as the 
mode with mode number $m=28$.
In contrast, the contribution below the resonance chain resembles the shape 
of the partial barrier associated to the $4:1$ resonance.
Note, that this contribution looks quite regular despite the absence of
regular tori in this phase space region.

\subsection{Perturbation Theory \label{sec:perturbation_th}}

In addition to the numerical solutions of the mode equation~\eqref{eq:mode_eq} also approximate
solutions can be obtained.
In particular the perturbative expansion of optical modes and wave numbers
$\kml$ in the deformation parameter $\epsilon$ has proven to yield good
agreement with
numerical results for sufficiently small deformation \cite{DubBogDjeLebSch2008}.
For this approach the modes and wave numbers $\tilde{k}_{m,l}$ of the circular
cavity provide the unperturbed basis.
For TM polarized light and angular mode number $m$ the
latter are computed as the roots of
\cite{DubBogDjeLebSch2008}
\begin{align}
S_m(\tilde{k}R) = n\frac{J^{\prime}_m(n\tilde{k}R)}
{J_m(n\tilde{k}R)}\
 - \frac{H^{\prime}_m(\tilde{k}R)}{H_m(\tilde{k}R)}.
\label{eq:modes_circle}
\end{align}
Here, $J_m$ and $J^{\prime}_m$ as well as $H_m$ and $H^{\prime}_m$ denote
Bessel and Hankel functions of the first kind and of order $m$ and their
derivative, respectively.
The radial mode number $l$
corresponds to the $l$-th root of $S_m$ when ordered by
ascending real part.
The wave numbers in second order perturbation theory for deformations of the
type of
Eq.~\eqref{eq:boundary} are given by \cite{DubBogDjeLebSch2008, KulWie2016b}
\begin{align}
k_{m,l} = & \tilde{k}_{m,l} - \frac{\epsilon^2}{4} \tilde{k}_{m,l}
\left(1 + 2 \tilde{k}_{m,l}
\frac{H^{\prime}_m}{H_m} \right)
\nonumber \\
 &-\frac{\epsilon^2}{4}(n^2 - 1)\tilde{k}_{m,l}^2\left(
 \frac{1}{S_{m + N}} +
 \frac{1}{S_{m - N}}\right),
 \label{eq:boundary_PT}
\end{align}
if $m>N$.
Here, $S_{m \pm N}$, $H_m$ and $H_m^{\prime}$ are evaluated at $\tilde{k}_{m,l}R$.
Only the last term contains the coupling of modes with different angular mode
numbers differing from $m$ by $\pm N$.
The contribution to the imaginary part
$\mathrm{Im\,}\delta \kml^{(\pm N)}$ caused by
these mode couplings is given by \cite{DubBogDjeLebSch2008}
\begin{align}
\mathrm{Im\,}\delta \kml^{(\pm N)} =
	-\epsilon^2
\frac{(n^2 - 1)\mathrm{Re \,}\tilde{k}_{m,l}}{2\pi|S_{m \pm N}H_{m \pm N}|^2}.
\end{align}
Therefore, in the perturbative framework enhancement of the negative imaginary
part may occur if $S_{m \pm N}(\tilde{k}_{m,l}R)$ becomes small.
This, however, corresponds to $\tilde{k}_{m,l}$ being almost degenerate with
a mode of angular quantum number $m \pm N$.
As we study modes with $l=0$ and
$\mathrm{Re\,}k_{m, 0} < \mathrm{Re\,}k_{m + N, 0}$ holds,
this degeneracy is only
possible between modes with mode numbers $(m, 0)$ and $(m - N, \tilde{l})$ for
some $\tilde{l}>0$.
Thus $\mathrm{Im\,}\delta \kml^{(-N)}$ may cause the peaks seen in
Fig.~\ref{fig:wave_numbers0} in the numerically obtained wave numbers.
There the perturbatively obtained wave numbers for
both the near-integrable and the mixed system are shown as
open magenta diamonds
in Fig.~\ref{fig:wave_numbers0}(a)~and~(b), respectively.
In both cases they give rise to the correct initial exponential decay
of $\ImRkml$.
In the near-integrable system perturbation theory accurately predicts the first
peak but fails to describe the second peak.
In contrast for the mixed system the perturbative description fails to describe
even the first peak.
This failure can be traced back to coupling between modes whose
angular mode numbers differ by $\pm N$.
In the following sections we will
argue that the relevant couplings occur between modes whose
angular mode numbers differ by multiples of the order $a=4$ of the relevant
$4:1$ resonance.
In the near-integrable system we have $a=N$ and thus second order
perturbation theory resolves the first peak.
In contrast in the mixed system we have $a = 2N$ and thus second order
perturbation theory fails to explain the first peak.
Additional couplings appear in higher orders of perturbation theory only.
Whether or not they give rise to additional peaks is so far an open question,
which will not be discussed here.

\section{Ressonance--Assisted Tunneling \label{sec:RAT}}

In the following we derive the framework which allows for a
perturbative as well as a semiclassical description of complex
wave numbers in optical microcavities.
To this end we construct a canonical transformation from Birkhoff
coordinates $(s,p)$ to adiabatic action--angle coordinates $(S, P)$.
In these coordinates the dynamics in the vicinity of the relevant
nonlinear resonance can be approximated by an integrable pendulum-like
Hamiltonian.
In addition the adiabatic action coordinates allow for EBK quantization of the
full system,
which establishes a connection between wave numbers $\kml$ and quantizing
adiabatic actions $\Pml$.
This correspondence enables us to use quantum perturbation theory and to
give an accurate description of decay rates under the influence of
resonance--assisted
tunneling, which extends the results from Ref.~\cite{KulWie2016b} to systems
with a mixed phase space.
However this approach is limited to cases where the cavity under consideration
can be
approximated by a classical integrable cavity.
Using a ray-based description of the decay of optical modes we also overcome
this limitation making the perturbation theory applicable for arbitrary smooth
deformations.
Furthermore, we apply semiclassical methods developed for quantum maps to
compute the coupling between whispering--gallery modes
and faster decaying modes mediated by the nonlinear resonance.
Using the ray-based model of decay for both contributing modes
allows for an
intuitive description of complex wave numbers of optical modes
based on classical properties.

\subsection{Adiabatic Action--Angle Coordinates \label{sec:CanonicalTrafo}}

\begin{figure}[]
\includegraphics[width=8.6cm]{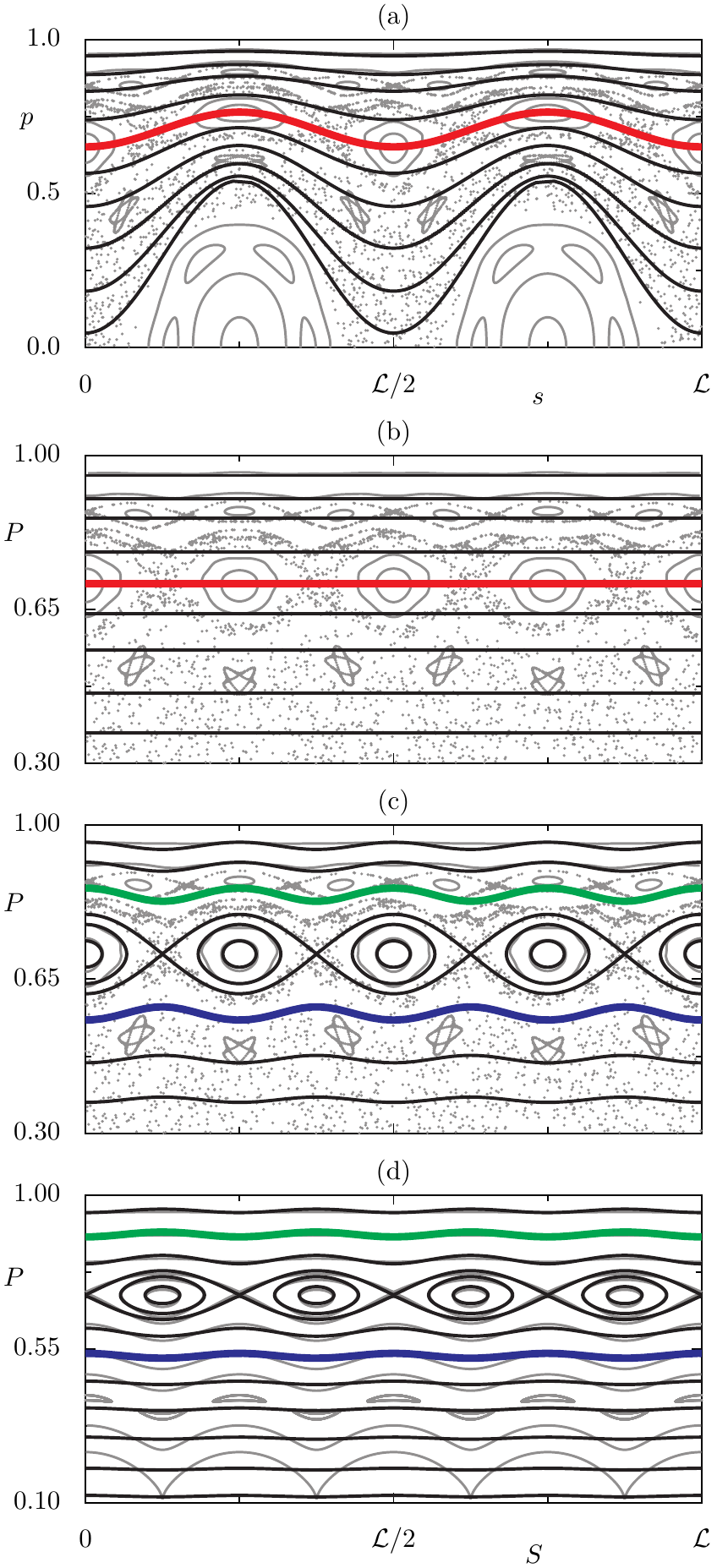}
\caption{\label{fig:adiabatic_invariants} In (a) the phase space of the mixed system (gray dots and lines) is superimposed by the adiabatically invariant curves in black in Birkhoff coordinates. The thick red line corresponds to the adiabatically invariant action $\Pab$. In (b) the same orbits and adiabatically invariant curves are shown in adiabatic action--angle coordinates. The phase space of the pendulum Hamiltonian is shown by black lines in (c) on top of the actual ray dynamics depicted in gray. The same scenario for the near-integrable system is shown in (d).  In (c) and (d) an example of the torus $\Pmlt$ (thick green line) and its symmetric counterpart $\Pratt$ (thick blue line) is shown.}
\end{figure}

The basis of our construction are adiabatically invariant
curves \cite{RobBer1985, NoeStoCheGroCha1996}
\begin{align}
p(s, P) = \sqrt{1 - \left( 1 - P^2\right)\left[R\kappa(s)\right]^{2/3}}
\label{eq:adiabatic_curve}
\end{align}
derived in the limit
of whispering gallery trajectories, i.e., $p \to 1$, and
parameterized by $P \in \left[0, 1\right]$, where $\kappa(s)$ denotes the
curvature of the boundary.
Here, we restrict the discussion to positive momenta.
The term adiabatically invariant is justified as the dynamics along the curve
is much faster than in the perpendicular direction.
The parameter $P$ describes the average momentum around which $p(s, P)$
oscillates.
For boundaries described by Eq.~\eqref{eq:boundary} the adiabatically invariant
curves Eq.~\eqref{eq:adiabatic_curve} can be expanded in $\epsilon$ to first
order if the polar angle $\phi$
is approximated by $2 \pi R s/ \mathcal{L}$.
This yields
\begin{align}
p(s, P) = P - \epsilon\frac{1- P^2}{3P}\left(N^2 - 1\right)
\cos\left(\frac{2\pi N}{\mathcal{L}}s\right).
\label{eq:adiabatic_curve_approx}
\end{align}
For several values of $P$ these adiabatic curves are superimposed on the
original phase space for the mixed system in
Fig.~\ref{fig:adiabatic_invariants}(a).
Note that on the one hand the adiabatically invariant curves smoothly
interpolate through nonlinear resonances.
On the other hand they provide a good approximation for  whispering--gallery
trajectories associated to the modes of interest in this paper.
Thus we wish to describe the dynamics in terms of $P$.
To this end we interpret Eq.~\eqref{eq:adiabatic_curve} as derivative of a type
two generating function $F_2(s, P)$ with respect to the arc--length coordinate
$s$.
We obtain this generating function by integrating
\begin{align}
F_2(s, P) & = \int_0^s p(s' , P)\mathrm{d}s'\label{eq:generatingfunc1} \\
& = sP - \epsilon\frac{1- P^2}{3P}
\frac{\mathcal{L}\left(N^2 - 1\right)}{2\pi N}
\sin\left(\frac{2\pi N}{\mathcal{L}}s\right)
\end{align}
and setting the undetermined $P$-dependent constant of integration zero.
The associated canonical transformation is then implicitly given by
Eq.~\eqref{eq:adiabatic_curve}~or~\eqref{eq:adiabatic_curve_approx} and
\begin{align}
S(s, P)  & = \frac{\partial F_2(s, P)}{\partial P} \\
& = s + \epsilon\frac{1 + P^2}{3P^2}
\frac{\mathcal{L}\left(N^2 - 1\right)}{2\pi N}\sin\left(\frac{2\pi
	N}{\mathcal{L}}s\right) , \label{eq:dF2_ds}
\end{align}
which has to be solved for
$(s, p) \in \left[0, \mathcal{L}\right) \times \left[0, 1\right]$ given $(S, P)
\in \left[0, \mathcal{L}\right) \times \left[0, 1\right]$ and vice versa. Note,
that the transformation is not global, as there is a minimal $P$ for which the
adiabatically invariant curve Eq.~\eqref{eq:adiabatic_curve} is real-valued for
all $s$.
The phase space of the mixed system in adiabatic action--angle coordinates is
depicted in Fig.~\ref{fig:adiabatic_invariants}(b) as dots superimposed by the
adiabatic curves $P = \mathrm{constant}$.
Geometrically, the canonical transformation straightens the overall curvature
of phase-space structures caused by the dynamics in the vicinity of the minimal
length periodic orbits at $p=0$.

\subsection{Pendulum Hamiltonian \label{sec:pend_ham}}

We continue with the construction of a pendulum Hamiltonian using the adiabatic action--angle coordinates introduced above.
Such a Hamiltonian is known to describe the dynamics in the vicinity of a nonlinear resonance and to capture the essential features of resonance--assisted tunneling \cite{BroSchUll2002}.
In the case of optical microcavities or billiards the Hamiltonian is given by \cite{Ozo1984, BroSchUll2002}
\begin{align}
H_{a:b}(S, P) = H_0(P) +
2V_{a:b}\cos\left(\frac{2\pi}{\mathcal{L}}aS + \phi_{a:b}\right),
\label{eq:pendHamiltonian}
\end{align}
where $H_0(P)$ as well as the parameters $\Vab$, and $\phi_{a:b}$ have to be
determined by the dynamics of the original system. \\

In particular $H_0(P)$ encodes the frequencies $\omega(P)$ of motion along
adiabatically invariant curves given by $P$.
Neglecting the effects of the boundary deformation these frequencies are given
by the frequencies of a circular cavity,
\begin{align}
\omega(P) = \frac{\mathcal{L}}{\pi}\arccos(P).
\label{eq:frequencies}
\end{align}
A more rigorous argument leading to this equation can be made using averaging
\cite{DulMei2012}
for the effective map
derived in Refs.~\cite{RobBer1985, Noe1997}.
In particular Eq.~\eqref{eq:frequencies}  guarantees the correct limit of
whispering-gallery
trajectories almost tangential to the boundary, i.e., $P=1$ corresponds to a
line of fixed points as $\omega(1) = 0$.
On the other hand, as $\omega(0) = \mathcal{L}/2$, for the trajectory
bouncing along the symmetry line of the billiard the arc-length coordinate $s$
advances by $\mathcal{L}/2$ between successive bounces.
In between these extreme cases Eq.~\eqref{eq:frequencies} follows the
frequencies of the circular cavity up to rescaling by
$\mathcal{L}/(2\pi R)$. \\

Given the frequency relation Eq.~\eqref{eq:frequencies}, an $a:b$ resonance
occurs for $\omega_{a:b} = b\mathcal{L}/a$, i.e., for
\begin{align}
\Pab = \cos\left(\pi \frac{b}{a}\right). \label{eq:P_ab}
\end{align}
This momentum $\Pab$ corresponds to the adiabatically invariant curve
which interpolates through the nonlinear resonance, i.e., on which the
periodic orbits are located.
For the relevant $a:b = 4:1$ resonance the adiabatic curve corresponding to
$\Pab$ is shown as thick red line in Fig.~\ref{fig:adiabatic_invariants}(a)
and (b).

In order to obtain the same frequencies as Eq.~\eqref{eq:frequencies} relative
to the resonant frequency we require $H_0^{\prime}(P) = \omega(P) -
\omega_{a:b}$, which yields \cite{KulWie2016b}
\begin{align}
H_0(P) = & \frac{\mathcal{L}}{\pi}\left(P \arccos(P) -
\sqrt{1 - P^2} + \sqrt{1 - \Pab^2} \right) \nonumber \\
& - \omega_{a:b}P.
\label{eq:H0full}
\end{align}
The quadratic expansion of Eq.~\eqref{eq:H0full} around $\Pab$ then leads to
\begin{align}
H_0(P) = \frac{1}{2 M_{a:b}}\left(P - P_{a:b}\right)^2
\label{eq:H0quad}
\end{align}
with
\begin{align}
\Mab = - \left( \frac{ 2 \pi}{\mathcal{L}}\right) \frac{1}{2}\sqrt{1 - \Pab^2}.
\label{eq:M_ab}
\end{align} \\

Having fixed $\Mab$ the parameter $\Vab$ controls the width of the
nonlinear resonance in phase space.
For near-integrabe systems it is related to the area $\Aab$ enclosed by the
separatrix of the unstable periodic points by \cite{EltSch2005}
\begin{align}
\Vab = \left(\frac{2\pi}{\mathcal{L}}\right)^2\frac{\Aab^2}{512 \Mab}.
\label{eq:VabArea}
\end{align}
Further away from integrability, e.g. for the considered mixed system, this
naturally generalizes to the area enclosed by the invariant manifolds of the
unstable periodic points up to their first heteroclinic intersection.
In contrast, for the near-integrable case $\Vab$ is determined by
the linearized dynamics around the stable periodic orbit in terms of its
monodromy matrix $\mathcal{M}_{a:b}$, which is known analytically for billiard
systems \cite{DulRicWit1996}.
This yields
\begin{align}
\Vab =  \left(\frac{\mathcal{L}}{2\pi}\right)^2
\frac{\Mab}{2a^4}\left[\arccos\left( \frac{\mathrm{Tr}
	\mathcal{M}_{a:b}}{2}\right) \right]^2
\label{eq:VabMonodromy}
\end{align}
and is used for the near-integrable system investigated in this paper. \\

Finally, $\phi_{a:b}$ is fixed by matching the $S$-coordinate of the equilibria
of the resonance in the Hamiltonian~\eqref{eq:pendHamiltonian} with the
periodic orbits of equal
stability. The phase space of the resulting Hamiltonian for the mixed system and for the near-integrable system is
shown in Fig.~\ref{fig:adiabatic_invariants}(c) and (d), respectively.
Note, that in the near-integrable system good agreement between ray dynamics and
pendulum
Hamiltonian is already achieved, if all $\epsilon$ dependent corrections are
neglected.
However, we keep this corrections to study both the near-integrable and mixed
system within the same framework.

\subsection{EBK Quantization of Adiabatic Actions \label{sec:ebk}}

As both the perturbative and the semiclassical description of
resonance--assisted tunneling crucially depend on the pendulum
Hamiltonian~\eqref{eq:pendHamiltonian} defined on the boundary
a connection between actual modes of the cavity and adiabatic invariants has to
be established.
In integrable systems this correspondence is given by means of EBK quantization of invariant tori
of the full ray dynamics.
In systems with two degrees of freedom the existence of such invariant tori implies the existence of a second conserved quantity besides energy.
For the integrable elliptic cavity, which includes circular cavities as special case,
this quantity is given by the adiabatic invariant $P$, i.e., $P$ is an actual invariant of the
dynamics for all times.
In contrast, for generic systems $P$ changes adiabatically.
However, if this change is slow enough, the corresponding adiabatically invariant, two-dimensional tori may still be used for EBK quantization.
However, good agreement with numerically determined wave numbers can only be expected, if the actual ray dynamics follows the adiabatic invariant curves for a sufficiently long time.
This assumption is expected to be fulfilled when these curves are located in the regular part of phase space.
In particular this is the case for whispering-gallery trajectories. \\

Following Refs.~\cite{KelRub1960,NoeSto1997,Noe1997} the 
real part of the wave number associated with a
quantizing, adiabatic invariant $\Pml$ 
corresponding to a whispering--gallery mode
with  angular mode number $m$ and radial mode
number $l$ is given by
\begin{align}
\Rekml = \frac{2 \pi m}{n}\left(\int_0^{\mathcal{L}}p\left(s, \Pml\right)\mathrm{d}s\right)^{-1}.
\label{eq:EBK_wavenumber}	
\end{align}
Here, the adiabatic invariant $\Pml$ is required to fulfill the 
quantization condition \cite{Noe1997}
\begin{align}
m \left(L(0, s^*) -  \int_0^{s^*}p\left(s, \Pml\right)\mathrm{d}s \right) \nonumber \\
= \left(l + \alpha \right) \int_0^{\mathcal{L}}p\left(s, \Pml\right)\mathrm{d}s,
\label{eq:EBK_quant}
\end{align}
where, $s^*$ denotes the arc--length coordinate of the first collision of a ray
started with initial conditions $(0, p(0, \Pml))$ on the boundary
and $L(0, s^*)$
denotes the geometric length of the ray segment in between the
boundary points labeled by $s=0$ and $s=s^*$, respectively.
The parameter $\alpha$ represents the phase shift that an incoming wave
undergoes when it is
reflected at the cavity boundary.
Thus $\alpha$ takes the openness of the system and the corresponding boundary
conditions
into account.
As there is no analytical result for $\alpha$ in the case of open cavities we
fix  its value at zero deformation where the wave numbers $\kml$ are known.
Specifically, in the circular case Eq.~\eqref{eq:EBK_wavenumber} reduces to
$p(s, \Pml)= \Pml = m/(n\mathrm{Re\,} \kml)$,
which allows to fix $\alpha$ in Eq.~\eqref{eq:EBK_quant}.
We further assume, that $\alpha$ remains constant under deformation of the
boundary.
This allows us to solve Eq.~\eqref{eq:EBK_quant} numerically for $\Pml$.
For this purpose we use the exact
expression~\eqref{eq:adiabatic_curve} for $p(s,\Pml)$.
Note, that Eq.~\eqref{eq:EBK_quant} does not necessarily permit a solution for
arbitrary $m$ and $l$.
For instance in case of the mixed system studied in this paper we
obtain $2l + 1 < m$ as a necessary condition for the existence of solutions.

\begin{figure}[h]
	\includegraphics[width=8.4cm]{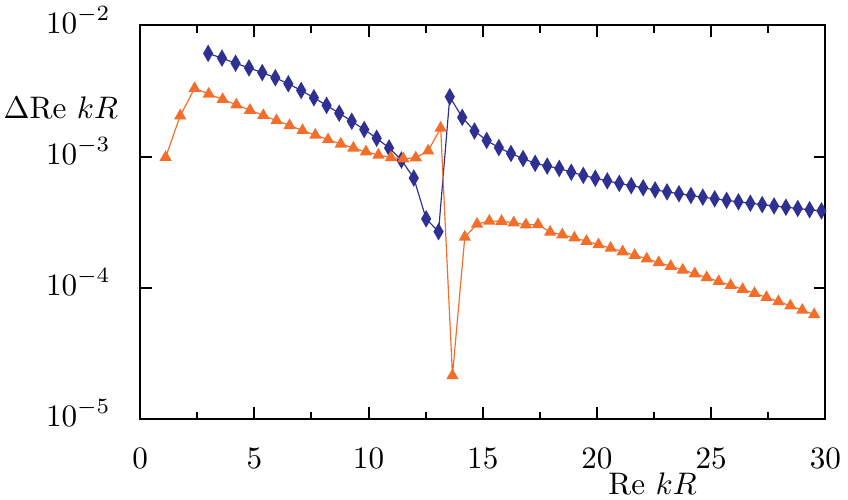}
	\caption{\label{fig:ebk_vs_bem}
		Relative errors of EBK wave numbers given by
		Eq.~\eqref{eq:EBK_vs_BEM} for the near-integrable system (orange
		triangles) and for the mixed system (blue diamonds). The lines
		connecting the symbols are a guide to the eye.}
\end{figure}

The accuracy of this approach has been demonstrated in Ref.~\cite{NoeSto1997}.
When applied to the near-integrable and the mixed system the quantization scheme
presented above yields good agreement with numerically computed wave numbers, as
Fig.~\ref{fig:ebk_vs_bem} shows.
There, the normalized error
\begin{align}
\Delta \Rekml = \frac{|\Rekml^{\mathrm{(EBK)}} -
	\Rekml^{\mathrm{(BEM)}}|}{\Rekml^{\mathrm{(BEM)}}}
\label{eq:EBK_vs_BEM}
\end{align}
between numerical (BEM) and semiclassical (EBK) obtained wave numbers is
depicted as a function of the real part of the wave number.
For both near-integrable system and mixed system, represented by the
orange triangles and blue diamonds, respectively, the error
becomes smaller with increasing real part of the wave number.
In both cases larger deviations occur around the wave number with $m=23$, where
the peak in the negative imaginary parts was observed.
This can be traced back to the fact that the EBK quantization scheme neglects
the coupling of modes associated with different quantizing adiabatic invariants
while such mode coupling may cause a shift in the real part of the wave number.
Overall, the relative error made by EBK quantization is slightly smaller in
the near-integrable system.

\subsection{Ray-Based Decay \label{sec:ray_decay}}

While EBK quantization yields accurate real parts of the wave numbers,
a ray-based quantitative description of their imaginary parts is
available only for fully chaotic systems or the special case of the integrable
circular cavity.
However, in the following we present a ray-based model of decay rates,
which yields good agreement with the numerically
obtained imaginary parts of wave numbers.

To this end, we assign an effective reflectivity $R(P)$ for a light ray moving on an
adiabatically invariant curve given by $P$.
This is accomplished by averaging
the reflectivity $R(p)$ of a circular cavity obtained in Ref.~\cite{HenSch2002}
along the adiabatically invariant curve defined by $P$ and
Eq.~\eqref{eq:adiabatic_curve_approx}.
That is, we set
\begin{align}
R(P) = \frac{1}{\mathcal{L}} \int_0^{\mathcal{L}} R(p(s, P))\frac{\partial S(s,P)}{\partial s}\mathrm{d}s
\label{eq:reflectivity}
\end{align}
where a uniform density along the adiabatic curve in the adiabatic action--angle
coordinates is assumed.
The corresponding density $\partial S/\partial s$ in Birkhoff coordinates is
obtained by differentiating Eq.~\eqref{eq:dF2_ds} with respect to the arc-length
coordinate.
Equating the exponential decay in the wave and in the ray picture gives the
decay rate \cite{HenSch2002}
\begin{align}
\Gamma = -2c \frac{\ln(R(P))}{4n R \sqrt{1 - P^2}}
\label{eq:Imk_raybased}
\end{align}
of a mode associated with $P$.
Note that $P$ is not required to fulfill a quantization
condition.
Thus it is applicable for arbitrary $P$.
In contrast, for quantizing $P=\Pml$ we have
\begin{align}
\Imkml = -\frac{\Gamma_{m,l}}{2c}.
\label{eq:Imk_raybased2}
\end{align}
Strictly speaking this is only true for integrable systems, where the actual
modes are associated with a single classical torus defined by $\Pml$.
We call $\Gamma_{m,l}$ the direct decay rate of this mode.
For nonintegrable systems dynamical tunneling between quantizing adiabatically
invariant tori will cause the decay rates and therefore also the imaginary
parts of wave numbers to deviate from the direct decay given by
Eq.~\eqref{eq:Imk_raybased}.

\subsection{Perturbation Theory of Resonance--Assisted Tunneling  \label{sec:rat_pert_th}}

While the ray-based model of decay describes the direct decay of an
optical mode associated with the quantizing adiabatic invariant,
resonance--assisted tunneling causes couplings to faster decaying modes.
As these couplings are small, the real part of the wave numbers is expected to
be not influenced by resonance--assisted tunneling, while the overall decay
will increase.
This enhancement is well described by quantum perturbation theory within
the Hamiltonian~\eqref{eq:pendHamiltonian}.
Based on this, we construct a perturbative expansion
of the imaginary parts of the wave numbers $\kml$ for whispering--gallery modes
with $l=0$.
For quantum perturbation theory an unperturbed basis in terms of optical modes
and wave numbers is required.
As mentioned above, this is accomplished either by the circular
cavity, which approximates the near-integrable system and the elliptic
cavity approximating the mixed system, respectively.
This allows for a
perturbative
treatment of resonance--assisted tunneling in both systems.
Moreover, we will demonstrate that using the EBK quantization scheme and the
ray-based model for the imaginary parts of wave numbers an
actual
approximating cavity is not necessary.
More specifically, we construct the perturbative expansion solely from the EBK
wave numbers, quantizing adiabatic invariants and the decay rates introduced in
Sec.~\ref{sec:ray_decay}. \\

In order to obtain a perturbative description we follow 
Refs.~\cite{LoeBaeKetSch2010, SchMouUll2011} and apply the
perturbation theory of resonance--assisted tunneling 
in quantum systems to optical microcavities.
To this end let us denote the wave numbers of the approximating cavity by
$\tilde{k}$ and expand the imaginary part of $\kml$ as
\cite{LoeBaeKetSch2010, KulWie2016b}
\begin{align}
\Imkml = \mathcal{N}^{-2}\sum_{
	v \geq 0} |c_v|^2  \mathrm{Im }\tilde{k}_{m - va,l + vb}.
\label{eq:RAT_PT}
\end{align}
Here, $\mathcal{N}^{2}=\sum_{v \geq 0} |c_v|^2$ accounts for proper
normalization and the sum is restricted by the requirement that all
angular mode numbers $m - va$ have to be non-negative.
Further note, that only wave numbers of modes within the same symmetry class
contribute to the sum.
The coefficients $c_v$ follow from
the perturbative scheme developed in
Ref.~\cite{Loe1951} applied to the pendulum
Hamiltonian~\eqref{eq:pendHamiltonian} and 
adapted to optical systems \cite{LoeBaeKetSch2010,YiKulKimWie2017}.
They are determined by the wave numbers of the approximating
cavity and the corresponding quantizing adiabatic invariants.
In order to compute the coefficients $c_v$ we make the following observations.
Given an approximating cavity its adiabatically
invariant curves, Eq.~\eqref{eq:adiabatic_curve}, should
coincide with
the adiabatically invariant curves of the original system.
Therefore, by means of EBK quantization, the quantizing adiabatic
invariants $P_{m - va,l + vb}$ coincide as well, if we further
assume the length
$L(0, s^*)$ of the ray segment appearing in
Eq.~\eqref{eq:EBK_quant} to be approximately the same.
Additionally this implies that also the real parts of the wave numbers
in both systems agree, i.e.,
$\mathrm{Re\,}\tilde{k}_{m - va,l + vb} = \mathrm{Re\,}k_{m - va,l + vb}$.
The coefficients $c_v$ can be computed by assigning an energy
\begin{align}
E_{m - va,l + vb} = H_0(P_{m  - va, l + vb} - \dpos(k_{m  - va, l + vb}))
\end{align}
in the Hamiltonian $H_0(P)$ given by
Eq.~\eqref{eq:H0full} to each of the quantizing adiabatic invariants.
Note, that here we take the periodic-orbit shift into account, which we assume
to be the same in both coordinate systems as they are transformed into each
other by a near identity transformation.
This adjusts the effective position of the nonlinear resonance at
$\Pab +\dpos(k_{m  - va, l + vb})$ relative to $P_{m  - va, l + vb}$.
However, shifting the quantizing adiabatic invariants rather than $\Pab$ allows
to use the same Hamiltonian for all wave numbers.
The coefficients in the perturbative expansion~\eqref{eq:RAT_PT} then can be
written as \cite{YiKulKimWie2017}
\begin{align}
c_v =
\prod_{u \leq v}\frac{\Vab(\mathrm{Re}\,k_{m,l})^2\text{e}^{\mathrm{i}\phi_{a:b}}}{E_{m,l}
	(\mathrm{Re}\,k_{m,l})^2
- E_{m  - ua, l + ub}(\mathrm{Re}\,k_{m  - ua, l + ub})^2},
\label{eq:RAT_coeffs_corrected}
\end{align}
where in the product $u$ is restricted to those values for which the
quantization condition Eq.~\eqref{eq:EBK_quant} permits a solution.
As the nonlinear resonance only couples modes whose wave numbers are similar,
i.e.,
$\mathrm{Re}\,k_{m,l} \approx \mathrm{Re}\,k_{m - va, l + vb}$, the wave
numbers in
Eq.~\eqref{eq:RAT_coeffs} cancel (up to small corrections) and $\dpos(k)$
can be
assumed to be constant for fixed $\Rekml$.
This additional approximation does not
alter the end result significantly and thus
we use
\begin{align}
E_{m - va,l + vb} = H_0(P_{m  - va, l + vb} - \dPos(k_{m,l}))
\end{align}
as well as \cite{KulWie2016b}
\begin{align}
c_v =
\prod_{u \leq v}\frac{\Vab \text{e}^{\mathrm{i}\phi_{a:b}}}{E_{m,l}
- E_{m  - ua, l + ub}}
\label{eq:RAT_coeffs}
\end{align}
in the following.
Further note, that so far for the evaluation of Eq.~\eqref{eq:RAT_coeffs} only
properties of the original system are required.
In particular, it is not required that an approximating cavity exists. \\

\begin{figure}[t]
\includegraphics[width=8.4cm]{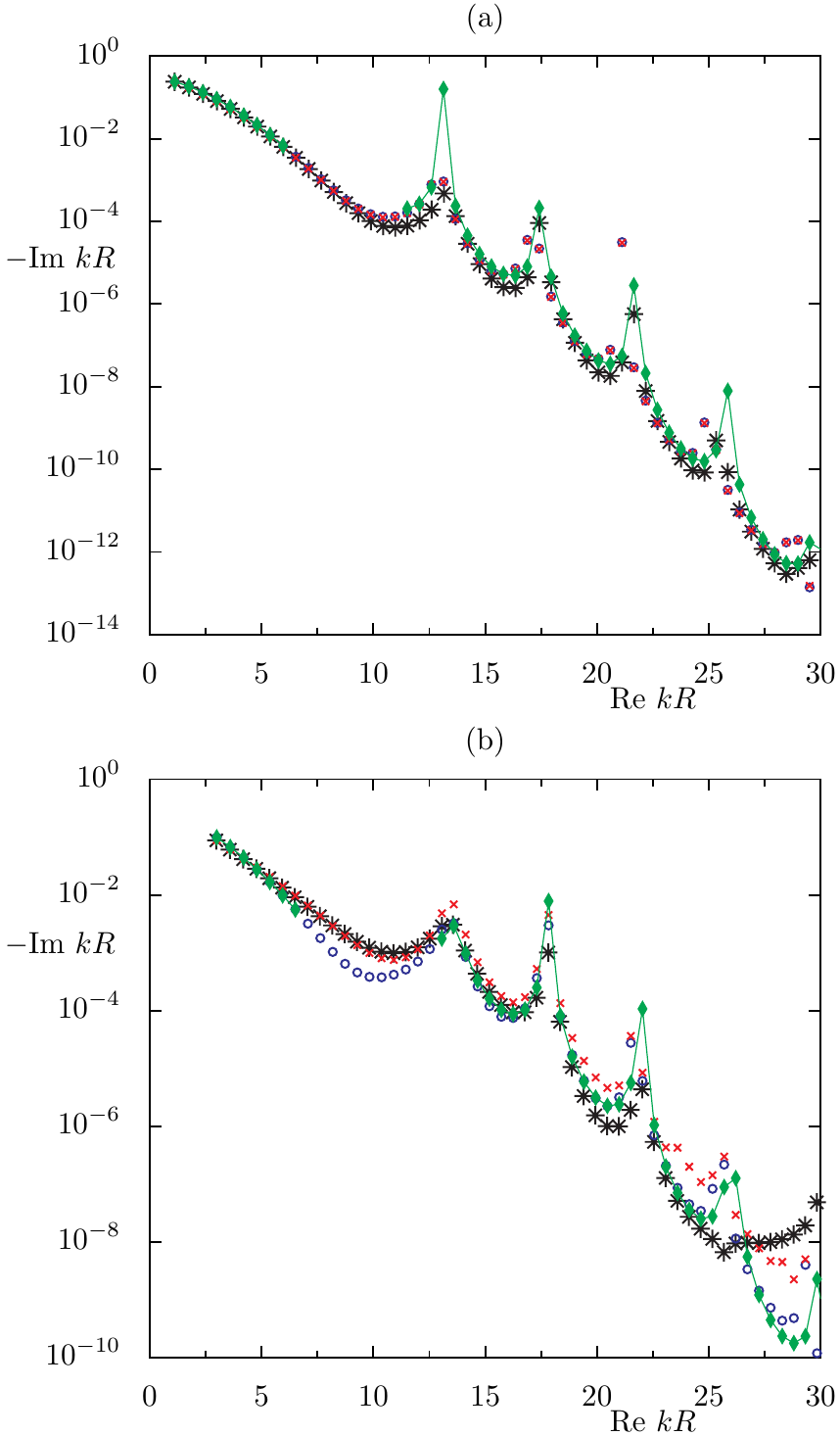}
\caption{\label{fig:wave_numbers} Wave numbers for the near-integrable systems
	are shown in (a) starting from $m=1$  and in (b) for mixed
	system starting from $m=4$. Numerical obtained wave numbers are
	depicted as black stars. The perturbative description using the imaginary
	parts of wave numbers in the circular and elliptic cavity is shown as red
	crosses while blue circles denote the perturbative description combined with
	the ray-based decay rates. The semiclassical description is represented by
	green diamonds.
	The lines connecting the data points are a guide to the eye.}
\end{figure}

Using the coefficients Eq.~\eqref{eq:RAT_coeffs} and the imaginary parts of
wave numbers in the circular and elliptic cavity we find overall good
qualitative agreement
with numerical obtained mode numbers in the near-integrable and the mixed
system.
This is shown in Fig.~\ref{fig:wave_numbers}(a)~and~(b)
where the numerically
obtained wave numbers (black stars)
are compared
with Eq.~\eqref{eq:RAT_PT} (red crosses).
For the near-integrable system the wave numbers of the
circular cavity can be obtained analytically.
For the mixed system the eccentricity of the elliptic cavity is chosen
according to
Eq.~\eqref{eq:eccentricity} and the wave numbers are computed numerically.
Thus for the mixed system the method provides no numerical advantage
compared to the direct numerical computation of wave numbers.
However it demonstrates the validity of the approach.
In both systems we truncate the perturbative expansion, Eq.~\eqref{eq:RAT_PT},
at $v\leq 5$.
In the near-integrable system
the exponential decay of
the negative imaginary part of the wave numbers match
with the numerically obtained wave numbers.
In contrast, in the mixed system the overall exponential decay matches up to
the second peak only, but is slower than in the numerical data for larger real
parts.
We assume this to be caused by the Goos--H{\"a}nchen shift rendering the
modified ray-dynamics of the elliptic cavity nonintegrable and the
corresponding enhancement of the decay of its optical modes, which are used as an
unperturbed basis.
In both systems the first numerically obtained peak is described
well.
The subsequent peaks are predicted systematically for slightly smaller real
parts of the wave numbers compared to the numerical data.
The strong enhancement of the negative imaginary part of the wave number
can be traced back to the coefficients Eq.~\eqref{eq:RAT_coeffs} which diverge
whenever a quantizing adiabatic invariant $P_{m  - va, l + vb}$ is energetically
degenerate with $\Pml$ with respect to $H_0$ when the periodic-orbit shift is
taken into account.
Within the quadratic approximation of $H_0$ by Eq.~\eqref{eq:H0quad} this
corresponds to $P_{m  + va, l - vb}$ and $\Pml$ being located symmetrically
around $\Pab + \dpos(\kml)$.
Deviations of the actual frequencies of motion along adiabatically invariant
curves from those predicted by $H_0$ may shift the wave numbers for which this
degeneracy occurs in the perturbative description.
Further, this degeneracy may be spoiled by deviations of the periodic orbit
shift from Eq.~\eqref{eq:dpos}.
In the mixed system the perturbative description fails when $\ReRkml > 25$ as
it does not include additional resonances which may cause the formation of the
observed plateau. \\

While the perturbative expansion gives
rise to qualitatively, and for the near-integrable system also quantitatively,
good agreement with numerical data,
Eq.~\eqref{eq:RAT_PT} still requires
the knowledge of the imaginary parts of the wave numbers of the approximating
integrable cavity.
Replacing them by Eq.~\eqref{eq:Imk_raybased2} and Eq.~\eqref{eq:Imk_raybased}
evaluated at $P_{m  - va, l + vb}$ allows
for a prediction of wave numbers without making reference to an actual
approximating cavity.
The resulting wave numbers are shown as blue circles in
Fig.~\ref{fig:wave_numbers}.
For the near-integrable system they perfectly match the prediction obtained
above.
In the case of the mixed system they underestimate the initial
exponential decay by up to one order of magnitude.
For larger real parts of the wave numbers they  agree slightly better with the
numerical data compared to the prediction using the wave numbers of the elliptic
cavity up to the wave numbers where additional resonances may become important.

Using this ray-based model of decay allows for an application
of perturbation theory when there is no actual approximating cavity.
The remaining inaccuracy of the ray-based model can be
associated with
the approximate nature of both the adiabatically invariant curves and the
associated effective reflectivity, Eq.~\eqref{eq:reflectivity}.
Moreover some of these adiabatically invariant curves
are located partially in the chaotic region of the actual ray dynamics.
Thus classical chaotic transport phenomena as well as chaos--assisted tunneling
may influence the decay.
These phenomena are not taken into account here.

\subsection{Semiclassical description \label{sec:RAT_sc}}

While the perturbative description of resonance--assisted tunneling
decomposes the imaginary part of the whispering gallery modes wave numbers into
several contributions from different modes, in a semiclassical description
there are only two contributions.
In particular there is a direct contribution to the decay denoted as
$\Gamma_{m,l}$ which
resembles the decay of an optical mode associated with the quantizing
adiabatic invariant $\Pml$.
Therefore, the direct contribution describes the decay of the mode that
would be present even if the system was integrable, i.e., in the absence
of nonlinear resonances and resonance--assisted tunneling.
The contribution resulting from resonance--assisted tunneling is denoted by
$\Gamma_{\mathrm{rat}}$ and is associated with an adiabatic
invariant $\Prat$ located symmetrically on the opposite side of the
nonlinear resonance with respect to $\Pml$.
In general $\Prat$ does not fulfill a quantization condition and thus is
not associated with an actual mode of the system or of an approximating
integrable cavity.
To be more precise, an optical mode associated with $\Pml$ obtained by EBK
quantization will also localize on $\Prat$.
However, the amplitude on $\Prat$ is suppressed by a factor $\AT$ called the
tunneling amplitude.
Taken this into account the imaginary parts of the wave numbers can be
decomposed as \cite{FriBaeKetMer2017}
\begin{align}
\mathrm{Im\,}k_{m,l} = -\frac{1}{2c}\Gamma_{m,l}
            - \frac{\AT^2 }{2c}\Gamma_{\mathrm{rat}}.
\label{eq:Im_k_sc}
\end{align}
Here, $\Gamma_{m,l}$ is given by
Eqs.~\eqref{eq:Imk_raybased}~and~\eqref{eq:Imk_raybased2} for $P=\Pml$
and $\Gamma_{\mathrm{rat}}$ for $P=\Prat$, respectively. \\

Thus it remains to compute $\Prat$ and $\AT$ in the following, which is done by
means of WKB theory within the pendulum
Hamiltonian~\eqref{eq:pendHamiltonian} with quadratic
$H_0(P)$,~Eq.~\eqref{eq:H0quad}.
Due to the periodic-orbit shift we consider $\Pmlt = \Pml - \dpos(\kml)$
for the semiclassic construction of $\Prat$ and $\AT$.
The equally shifted $\Pratt = \Prat - \dpos(\kml)$ is defined as the torus
on the opposite side of the resonance with
respect to $\Pmlt$ but with the same energy $E_{m,l} = H_0(\Pmlt)$.
It is given by $\Pratt = 2\Pab - \Pmlt$.
Both $\Pmlt$ and $\Pratt$ are shown schematically in
Fig.~\ref{fig:adiabatic_invariants}(c)~and~(d) for the mixed and the near
integrable system as thick green and blue lines, respectively.
In order to compute the tunneling amplitude we identify $1/(n \ReRkml)$ as the
semiclassical parameter which plays the role of the reduced Planck constant.
Using this identification gives
\begin{align}
\AT = \left|2\sin \left(\frac{n}{2a}\ReRkml\,\mathcal{A}_{\mathrm{rat}} \right) \right|^{-1}\exp{\left(-n\ReRkml\,\sigma\right)}
\label{eq:AT}
\end{align}
for optical microcavities.
Here,
\begin{align}
\mathcal{A}_{\mathrm{rat}} = \mathcal{L}(\Pml - \Prat)
\label{eq:Arat}
\end{align}
is the phase-space area bounded by the adiabatic invariants
$\Pml$ and $\Prat$ and
$\sigma$ is determined by the imaginary action of complex classical
paths
which bridge the nonlinear resonance and connect  $\Pmlt$ with $\Pratt$.
In particular, for the action of these complex paths we have
\begin{align}
\sigma = \frac{\mathcal{A}_{\mathrm{rat}}}{2\pi a}
	\ln\left(\left| \frac{4E_{m,l}}{\text{e}^2\Vab}\right|\right)
\label{eq:sigma}
\end{align}
in analogy to Ref.~\cite{BroSchUll2002}.
In Fig.~\ref{fig:wave_numbers} we compare the wave numbers obtained by the
semiclassical formula Eq.~\eqref{eq:Im_k_sc} with the numerical data.
Note, that the semiclassical description is only valid if the quantizing
adiabatic invariant is located outside of the $4:1$ resonance.
Hence the initial exponential decay in both the near-integrable and the mixed
system is
due to the direct contribution in Eq.~\eqref{eq:Im_k_sc} as $\Pmlt$ is located
below the resonance while $\Pratt$ is located above.
Therefore the direct contribution is dominant and the
resonance--assisted contribution can safely be neglected.
Similar to the perturbative description the initial decay is underestimated
in the mixed system as Eq.~\eqref{eq:Imk_raybased} underestimates the actual
decay within this regime.
In between the initial decay and the first peak the quantizing adiabatic
invariant $\Pmlt$ is located inside the $4:1$ resonance and thus no meaningful
semiclassical prediction based on Eq.~\eqref{eq:Im_k_sc} is possible.
Once $\Pmlt$ is located above the nonlinear resonance for larger real parts
of wave numbers the semiclassical description is again applicable and leads to good agreement with the numerically obtained wave numbers.
In particular, in
this regime the resonance--assisted contribution dominates and the
direct contribution can be neglected.
In both systems the overall exponential decay
is in good agreement with the numerical obtained wave numbers.
Moreover, the position of the peaks is resolved correctly in both
systems except for the peak around $\ReRkml = 25$ in the near-integrable system.
The peaks occur whenever the prefactor that enters in Eq.~\eqref{eq:AT}
diverges, which occurs if $n\ReRkml\,\mathcal{A}_{\mathrm{rat}}$ is an integer
multiple of $2\pi a$.
If this condition is fulfilled exactly, Eq.~\eqref{eq:AT} is ill-defined,
which causes the semiclassical prediction to overestimate the numerical
obtained negative imaginary parts.
At these wave numbers $\Prat$ satisfies a quantization condition as well.
That is $\Prat = P_{m  - va, l + vb}$ for some integer $v$.
As, by definition, $\Pml$ and $\Prat$ are energetically degenerate this is the
same condition obtained from the perturbative treatment of resonance--assisted
tunneling.
Again, for the mixed system, the semiclassical prediction does not capture the
plateau formation for wave numbers $\ReRkml > 25$, which we expect to be due to
additional resonances.

Although being at least equally accurate as the perturbative description the
semiclassical picture is also subject to the same errors induced by the
ray-based model of decay as discussed in the previous section. \\

\section{summary and outlook \label{sec:sum_outlook}}

In this paper we demonstrate how resonance--assisted tunneling gives rise to
enhanced decay of optical modes in deformed microdisks with near-integrable or
mixed phase space.
While the near-integrable case was treated before we extend the existing
perturbative description to systems far from integrabillity and presented also
a semiclassical description.
We apply both the perturbative and the semiclassical description to systems
with either near-integrable or mixed classical ray dynamics and find good
agreement with numerically obtained data for whispering--gallery modes.
In particular our description correctly captures the overall exponential decay
of the imaginary parts of wave numbers towards larger wave numbers and
predicts the wave numbers of modes with significantly enhanced decay.
In the latter case the semiclassical description gives rise to better agreement
with the numerical obtained positions of these peaks compared to perturbation
theory.
Moreover, we show that the theory of resonance--assisted tunneling predicts the
enhancement of decay in cases where the second order
perturbative expansion in the
deformation parameter does not apply.

Our approach is based on the construction of
adiabatically invariant action--angle coordinates
and the approximation of the relevant nonlinear resonance
by a suitable pendulum Hamiltonian.
We further use EBK quantization of adiabatic invariants and a ray-based model
for the decay of optical modes to compute wave numbers in the absence of
dynamical tunneling.
The obtained imaginary parts of the wave numbers can be interpreted as
the direct decay of the associated modes.
Subsequently, the pendulum Hamiltonian allows for the inclusion of
resonance--assisted tunneling by either quantum perturbation theory or a
semiclassical description.
Both descriptions confirm that the underlying mechanism of
enhancement coincides with what is known for two-dimensional quantum maps.
That is, the enhancement is due to the coupling of optical modes
associated with quantizing adiabatic invariants which are
located symmetrical with respect to the relevant nonlinear resonance.
This allows whispering gallery modes with slow decay to couple to faster
decaying modes, which can be seen also in their Husimi representation.
While the coupling between modes is well described within the pendulum
Hamiltonian the main
error of the presented approach is introduced by describing the direct decay
based on a simple ray picture.

In the regime of large wave numbers we expect also smaller resonances to become
important for tunneling as the numerical obtained wave numbers in the mixed system suggest.
These multi-resonance effects could be incorporated in the
perturbative description, as it was done for quantum maps.
However, a semiclassical picture thereof does not exist so far.

Furthermore, an obvious generalization of the deformed microdisks are
three-dimensional deformed spherical cavities, which are expected to exhibit
resonance--assisted tunneling as well.
As such cavities are frequently used in experiments an equally
good description
as in the two-dimensional case is of great interest.
However, with more degrees of freedom resonances of higher rank may arise which
give rise to a much
more complex way of resonance--assisted coupling between optical modes.
This was recently demonstrated for a normal--form
Hamiltonian \cite{FirFriKetBae2019}.
On an even more fundamental level the interplay of tunneling and classical
transport mechanisms, e.g., the famous Arnold diffusion, could become important
in  three-dimensional cavities.
This is, however, even an open problem for much simpler systems like
four-dimensional symplectic maps.

\begin{acknowledgments}
	We are grateful for discussions with
	Normann Mertig and Julius Kullig.
	Furthermore, we acknowledge support by the Deutsche Forschungsgemeinschaft
	under grant BA~1973/4--1.
\end{acknowledgments}

\end{document}